# Image Denoising for Strong Gaussian Noises With Specialized CNNs for Different Frequency Components

Seyed Mohsen Hosseini, smhosseini741@gmail.com

*Abstract*— in machine learning approach to image denoising a network is trained to recover a clean image from a noisy one. In this paper a novel structure is proposed based on training multiple specialized networks as opposed to existing structures that are base on a single network. The proposed model is an alternative for training a very deep network to avoid issues like vanishing or exploding gradient. By dividing a very deep network into two smaller networks the same number of learnable parameters will be available, but two smaller networks should be trained which are easier to train. Over smoothing and waxy artifacts are major problems with existing methods; because the network tries to keep the Mean Square Error (MSE) low for general structures and details, which leads to overlooking of details. This problem is more severe in the presence of strong noise. To reduce this problem, in the proposed structure, the image is decomposed into its low and high frequency components and each component is used to train a separate denoising convolutional neural network. One network is specialized to reconstruct the general structure of the image and the other one is specialized to reconstruct the details. Results of the proposed method show higher peak signal to noise ratio (PSNR), and structural similarity index (SSIM) compared to a popular state of the art denoising method in the presence of strong noises.

*Index Terms*— Image Denoising, Convolutional Neural Networks.

## I. INTRODUCTION

Image denoising aims to recover a clean image from a noisy one. Thermal noise and the noise created due to defects in image capturing devices are modeled by Additive White Gaussian Noise (AWGN). Removing AWGN, which is the aim of the proposed method here, is the subject of many literatures in this field. Recent developments in machine learning have created an incentive to use machine learning for image denoising. The model proposed in [1] is an effective machine learning denoising method. The network which is trained using residual learning, maps the noisy images to noise. The predicted noise then gets removed from the noisy image and the estimated clean image is created.

Another area where image denoising has been applied is the medical imaging field. In this field the image degradation resulting from limitations in image acquiring methods is treated as noise. For example machine learning is used to improve the quality of low dose computed tomography (LDCT) images [2],[3]. LDCT usually is used to reduce the dose received by patients but the image quality is lower than a full dose CT.

In most of the existing machine learning denoising methods the goal is to minimize the mean square error (MSE) over a collection of pixels. In these methods the average of errors is used as a metric to train the network and the recovery of the fine details of the image are overlooked, because the fine details do not impact the average error so much. So the MSE may be low but the visual integrity of the image is compromised. As a result the recovered image has a problem of over-smoothing and waxy artifacts. These artifacts are also present in state of the art model based denoising methods like BM3D [4]. This problem is especially serious in sensitive applications like medical imaging where the fine details and edges are important. Some approaches have been proposed to address this problem. For example in [5] an edge aware loss function is used to increase the accuracy of denoising in edge pixels to improve the perceptual quality of the recovered image. They propose a blind (not for a specific noise level) denoising network with two edge aware loss functions. First one is based on gradient magnitude from Sobel filter and the second one is a binary edge mask by thresholding the first one. They report a structural similarity measure (SSIM) improvement of 0.005 and 0.01, for the first and second edge aware loss functions. In [6] a generative adversarial network (GAN) with a visual attention mechanism to preserve the details is proposed. Their model consists of an attentive block, a generator and a discriminator. The attentive block generates an estimated map of noise distribution in the input image. This map is utilized in generator and decimator to help them focus on the noisy regions of the image.

Here separating low and high frequency information of the image is proposed to preserve details of the denoised image in the present of strong noise. So instead of one network trying to recover both general structures and details (low and high frequencies) specialized networks are trained to do so. Two different networks are trained for low and high frequency components of the image. One network is trained using the low frequency components of images so it will reconstruct general structures. The second one is trained using high frequency components of images and it will reconstruct the details. Separating general structure from details will prevent the details from being overlooked and will result in better image quality as the experimental results show.

## II. A DENOISING CONVOLUTIONAL NEURAL NETWORK, DNCNN

A lot of different convolutional neural network architectures have been proposed for image denoising. Among them the model proposed in [1] is a relatively simple structure with the state of the art performance in many cases. The architecture, for denoising of Gaussian noise with a specific noise level (DnCNN-S), consists of 17 convolutional layers, ReLU layers as activation functions, and batch normalization layers. Fig. 1 shows the architecture of the model. This network is trained with $128 \times 1600$ patches of size $40 \times 40$, and residual learning is utilizes in training. In [1] the network that is trained for a specific noise level is called DnCNN-S.

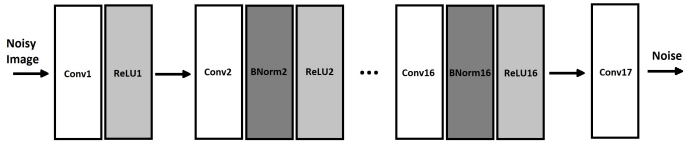

Fig. 1. The architecture of DnCNN [1].

The network learns the latent clean image in the input and removes it to produce the noise. So, instead of mapping noisy images to clean images, the network maps noisy images to the noise, which then will be removed from the noisy image. This method is a type of residual learning; know as global residual learning (GRL), as opposed to local residual learning (LRL) which, for example, is employed in ResNet [7]. Fig. 2 depicts the general structure of GRL and LRL. In LRL the input and output of every couple of layers are added, forming a ResBlock. Residual learning is an important technique in training of deep networks as it facilitates the flow of information and gradient.

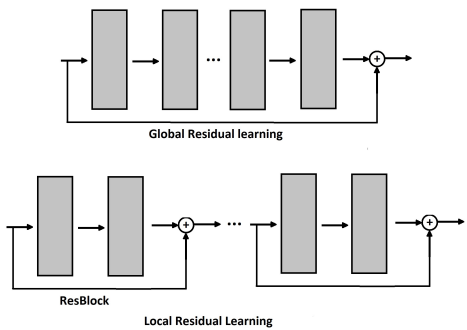

Fig. 2. Global and Local residual learning.

Using both batch normalization and residual learning is the main characteristics of DnCNN. The combination of these two methods has led to faster training and higher performances. A single model can be trained to handle different noise levels and different types of image degradation. DnCNN-3 which is a network with depth of 20 layers, and with filters of size $3 \times 3$, can reduce three types of noises: 1- Gaussian noise with specific noise levels from 0 to 55, 2- down-sampling degradation with down-sampling factors from 2,3,4, and 3- JPEG degradation with quality factor from 5 to 99. This network is trained with $128 \times 8000$ patches of size $50 \times 50$.

## III. THEORY AND IMPLEMENTATION OF THE PROPOSED MODEL

The main idea behind the proposed model is to separate the low frequency and high frequency information of the image and train a specific network for each of them. The popular denoising methods try to recover general structures and details with a single architecture. This leads to overlooking of details and producing over smoothed images.

**Advantages of the proposed model:** First, training two separate networks will provide twice as much learnable parameters, which will lead to capturing more information about the image and noise. For example, a 20 layer DnCNN has around 0.74 million parameters, and with two of them we will have around 1.48 million available learnable parameters. To have the same amount of learnable parameters with one network, a 40 layer network is needed. Training such a deep network will have the challenges of training very deep networks, like vanishing or exploding gradient. Special techniques and architectures have been proposed to solve the problems of training a very deep network. A successful model is ResNet [7].

The proposed model provides another technique for training a very deep network, by dividing it to two separate smaller networks, which are easier to train and can be trained separately and in parallel. For example, we would have the same amount of learnable parameters of a 50 layer network, with two separate 25 layer networks. Parallel training of the two networks is another advantage of the proposed model, which makes the training faster and more efficient.

Another advantage is the ability to change each network to make it more suitable for different image components. The number of layers, filter size, patch size, mini-batch size or other hyper-parameters can be chosen specifically for each network. So even if the training of a very deep network was not an issue, two specialized networks provide the advantage of having specialized hyper-parameters. Some parameters like patch size cannot be combined into one network, for example having a network with two different patch sizes. But some parameters can be combined in one very deep network, like a network with different filter sizes. The problem then becomes the flow of different types of information in a very deep network. The first layers of a convolutional network extract the general structures of an image and the deeper layers extract finer details. The general structure information then has to be transferred to deeper layers to reach the output layer alongside the finer details information. This process in very deep networks is not very efficient. In contrast in the proposed model, each of the two specialized networks has much less layers; this makes the flow of information easier and therefore the structure more advantageous than a very deep network.

Another main feature of the proposed model is not using residual learning. The proposed model is designed for very strong noises and residual learning is not very effective in this case. As it is stated in [1] when the noise is low the noisy image and the clean image are very similar. Mapping the noisy image to clean image in this case is more like an identity mapping, and optimizing a network for identity mapping is not very effective. But noisy image is very different to the noise and mapping the noisy image to noise (residual learning) is more successful.

In case of strong noises the situation is reversed. The noisy images are similar to noise and are different to clean image. So, mapping the noisy image to noise (residual learning) would be close to identity mapping, and the optimization would not be very successful. This is why residual learning has not been utilized in the proposed model, and the network maps noisy images to clean images. Without residual learning, in strong noises, better results have been observed compared to a model with residual leaning.

The structure of the proposed model consists of two CNNs similar to DnCNN-S, but it differs with DnCNN-S in a number of aspects; such as, different patch and filter sizes, different number of layers, and also training without residual learning. The networks are trained with noisy images as their input and low and high frequency components of original images as their target. The goal is denoising of the noisy image for a specific noise level, in future works the proposed model will be investigated in the present of different types of noise.

Fig. 3 shows the process of obtaining the low and high frequency components of original images. A Gaussian filter is applied to each image to create the low frequency component then that image is subtracted from the original image to obtain the high frequency component. Different values for the standard deviation of the Gaussian filter were tested and overall the standard deviation of 4.5 produced the best results with different levels of noise. With this value the information of the image is divided equally between two networks.

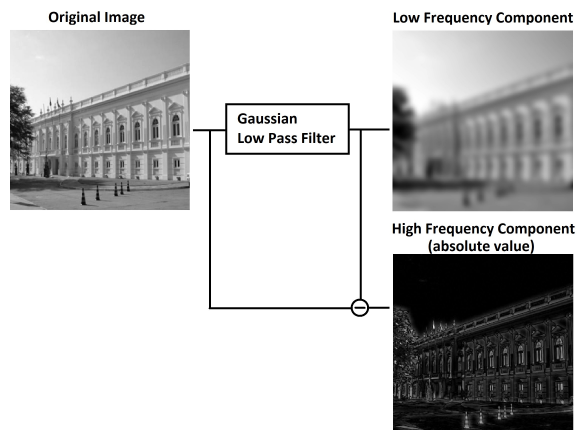

Fig. 3. Generating low and high frequency components of an image. This process is repeated for all of the training and test images.

The architecture of the proposed method is depicted in Fig. 4. The input patches are chosen randomly from the noisy images and the target patches are the corresponding patches from low and high frequency components of original images. The input patches of the two networks are the same but the target patches are different.

The patch size determines the amount of information that is available to the network, and how much information the network is able to capture is determined by its depth. So, the patch size is chosen in relationship with the receptive field of a network, which is defined by its depth. But for images with strong noise larger patch sizes will help capture more contextual information and thus produce better results [8]. The patch size for training of DnCNN-S is chosen in [1] as 40 × 40, based on the noise level of 25. Since the proposed model is designed for strong noises, the patch sizes are chosen larger than the receptive fields of the two networks.

The low frequency network has 17 convolutional layers, 1 layer with filter size 5 × 5 × 64, 15 layers with filter size 3 × 3 × 64, and 3 × 3 × 1 is the filter size of the last convolutional layer. 5 × 5 filter is used to capture general structure information more efficiently. There is batch normalization after each convolutional layer except the first and last one. ReLU layers follow all of the batch normalizations except the first layer where it follows a convolutional layer. For this network the patch size is 50 × 50 to make the network more suitable for strong noises.

High frequency network has 20 layers all of the filters are 3 × 3 × 64. The arrangement of layers is the same as the low frequency network. The patch size is chosen as 70 × 70. This patch size is larger than the patch size of low frequency network. Because there is more information in patches of high frequency component compared to patches of low frequency component. Increasing the patch size beyond this value requires a deeper network to capture all the available data. To keep the computational cost of the model down, the mentioned patch size and depth are chosen.

The model is trained for every noise level separately. Each network will be specialized to learn different information about the images. One will learn to map a noisy patch to low frequency component and the other one will learn to map the same noisy patch to the high frequency component. By adding the low and high frequency component the clean or denoised patch will be obtained. Recovering a clean patch from the noisy one using the trained specialized networks is depicted in Figure 5.

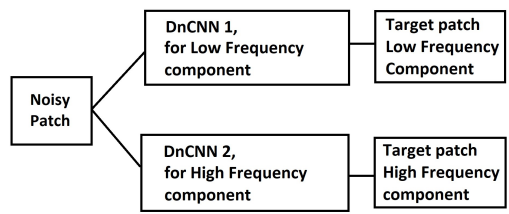

Fig. 4. the architecture of the proposed model. Two DnCNNs are trained using input patches form the noisy images and patches from low and high frequency components of the corresponding original images.

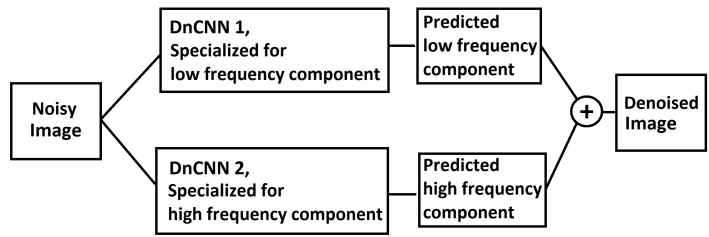

Fig. 5. denoising process using the proposed model.

## IV. EXPERIMENTAL RESULTS

The training data set is 545 images from IAPR TC-12 data set. The performance of the network is evaluated using the 200 test images from the BSDS500 dataset. The two networks of the proposed model were trained, with mini-batch size of 128 and 64 for low and high frequency networks respectively. 90000 image patches for each network (180000 in total). The learning rate was reduced for 0.1 to 0.0001 over 15 epochs (because of limited computing power). For comparison the results of two other models are also obtained. To keep the conditions similar, the training and testing of the 17 layer DnCNN-S was the same as training and testing of the proposed method (same number of patches, same number of epochs, same training and testing data sets). The other network is a DnCNN model that is included in MATLAB. This network is pre-trained for denoising of different types of noise and not for a specific Gaussian noise level. No training was done for this network. This network is included just to compare the performance of models trained for a specific noise level vs. a blind model (a model for various noise levels and noise types). Table 1 shows the average PSNR and SSIM for 200 test images from BSDS500 dataset, using the proposed method, a 17 layer DnCNN-S [1], and the blind DnCNN from MATLAB.

The results show that the proposed model in all noise levels out performs the state of the art DnCNN-S model. It can be concluded that when noise is strong the information is too much to be learned by one network. With strong noise the specialized network structure of the proposed model will give better results with higher PSNR and SSIM. The difference between PSNR and SSIM of the proposed model and DnCNN-S generally increases as the noise gets stronger. The advantage of the proposed model can be clearly seen in stronger noises. This better performance in strong noise levels is also partially due to not using residual learning. As it is stated in section 3, residual learning is more effective in relatively lower noise levels. Figures 6,7,8 show a number of noisy, original and denoised images obtained by DnCNN-S and the proposed model for noise levels 100 and 150. The two gray scale images are commonly used for method evaluation, and the color image is from BSDS500. The improvement of the waxy artifacts and over smoothing, like sharper edges, can be seen in the results of the proposed model.

TABLE 1

The average PSNR (dB) and SSIM values resulted from applying three models on 200 training images from BSDS500 dataset. The pre-trained DnCNN included in MATLAB (to show the effect of blind vs. non-blind denoising). A DnCNN-S [1] trained with the same conditions as the proposed model, except for residual learning, and the proposed model. The best results are highlighted in bold.

| Gaussian noise level | Blind DnCNN included in MATLAB | DnCNN-S | Proposed Specialized DN-CNNs | Difference (Proposed model - DnCNN-S) |
|---|---|---|---|---|
| 25 | 28.40 | 28.92 | **28.96** | +0.04 |
|    | 0.7944 | 0.8206 | **0.8255** | +0.0049 |
| 50 | 24.87 | 25.70 | **25.92** | +0.22 |
|    | 0.6562 | 0.7010 | **0.7192** | +0.0182 |
| 75 | 22.20 | 24.09 | **24.30** | +0.21 |
|    | 0.5320 | 0.6320 | **0.6535** | +0.0215 |
| 100 | 19.89 | 22.92 | **23.23** | +0.31 |
|     | 0.4121 | 0.5810 | **0.6096** | +0.0286 |
| 150 | 16.81 | 21.40 | **21.82** | +0.42 |
|     | 0.2629 | 0.5228 | **0.5587** | +0.0359 |

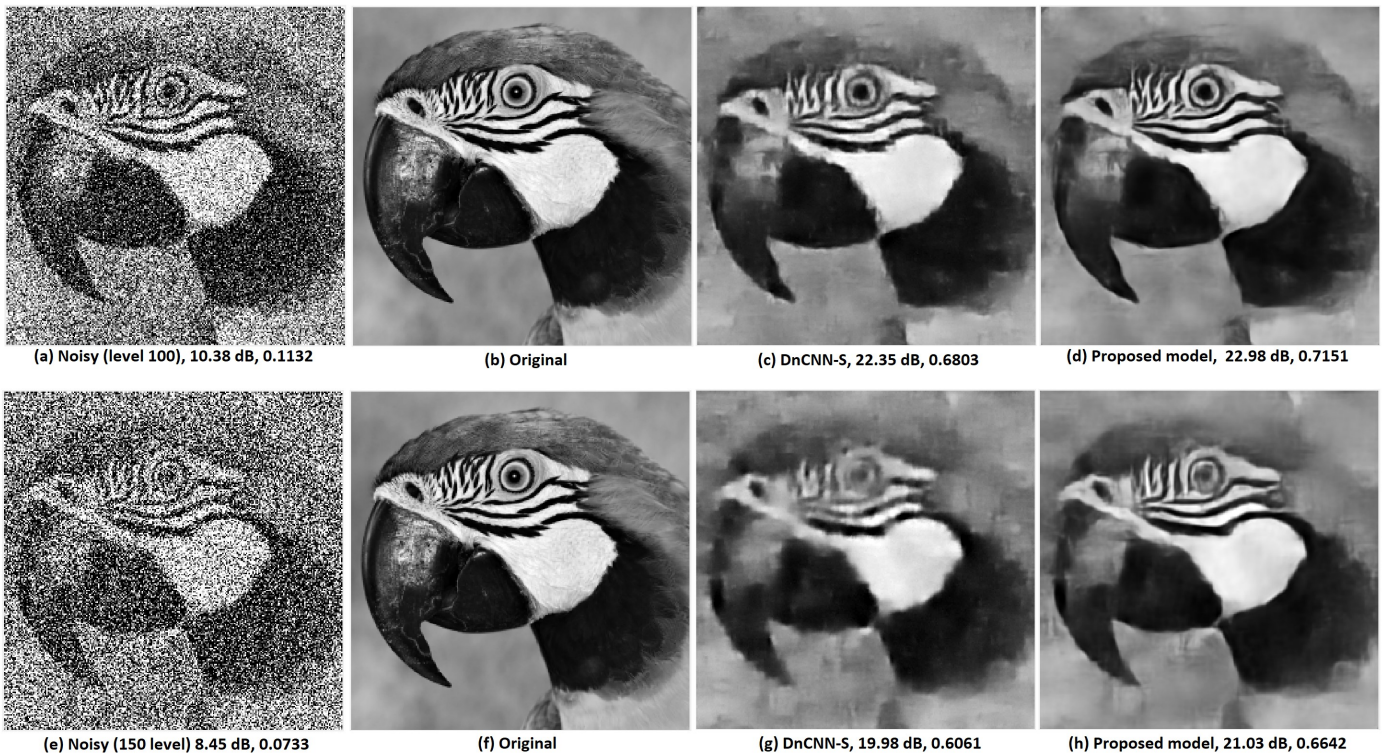

(a) Noisy (level 100), 10.38 dB, 0.1132　(b) Original　(c) DnCNN-S, 22.35 dB, 0.6803　(d) Proposed model, 22.98 dB, 0.7151

(e) Noisy (150 level) 8.45 dB, 0.0733　(f) Original　(g) DnCNN-S, 19.98 dB, 0.6061　(h) Proposed model, 21.03 dB, 0.6642

Fig. 6. Noisy, original and denoised images obtained by DnCNN-S and the proposed model, with PSNR and SSIM values, for noise levels 100 and 150.

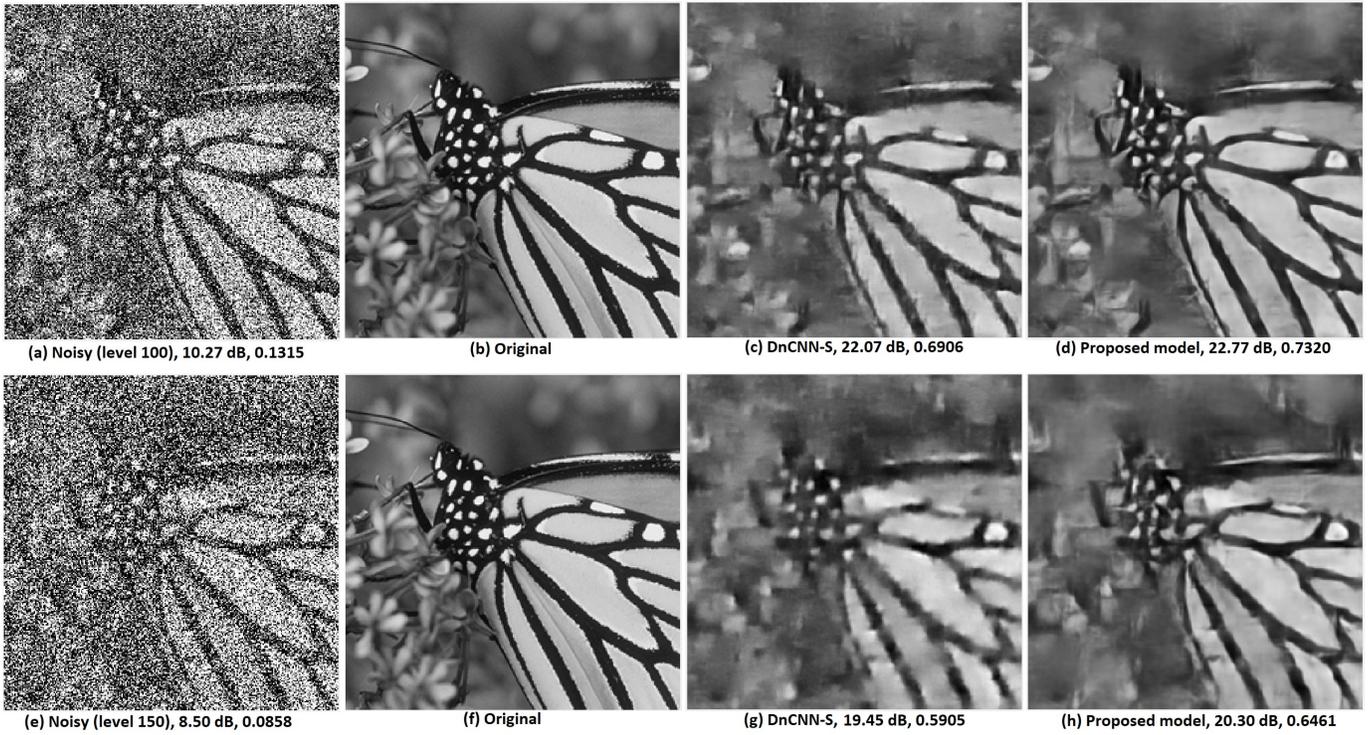

Figure 7 Noisy, original and denoised images obtained by DnCNN-S and the proposed model, with PSNR and SSIM values, for noise levels 100 and 150.

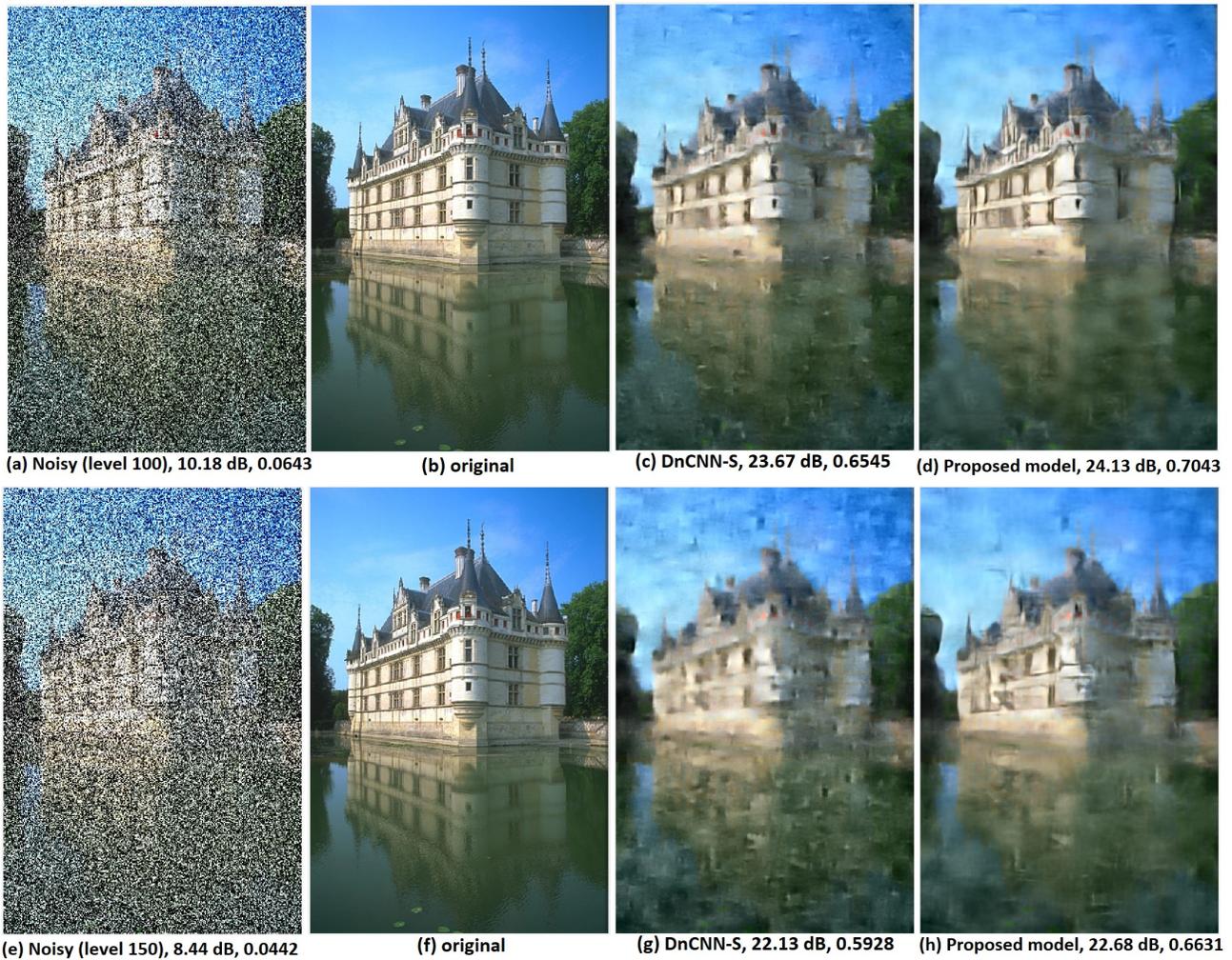

Figure 8 Noisy, original and denoised images obtained by DnCNN-S and the proposed model, with PSNR and SSIM values, for noise levels 100 and 150. Image form BSDS500.

## V. CONCLUSION AND FUTURE WORKS

A novel structure for denoising of known noise levels was proposed in this paper. The structure is based on separating the low and high frequency components of the target images and training two specialized networks. The proposed model is an alternative to training a very deep network. By dividing a deep network into smaller networks the issues of training a deep network, like vanishing or exploding gradient, can be avoided. The results show that in the presence of strong noise the proposed model outperforms the state of the art DnCNN-S model. The advantage of the proposed model becomes more pronounced as the noise increases. The connection between noise level and success of the proposed model indicates that the information in higher noise levels cannot efficiently be learned with one network, and using multiple specialized networks can improve the performance. In future works the performance of the proposed model for denoising of unknown noise levels and different types of noise will be investigated. The success of the proposed model in strong noise indicates that this model will perform better compared to a single network in the presence of different noise levels and noise types. Here a two network structure was investigated, but this model can be extended to multiple networks; for example: three specialized networks for low, middle, and high frequency components. A multiple network structure would have the ability of learning even more information about strong unknown noises of different types compared to popular one network structures. With this structure the learnable parameters of a very deep network would be available but we have to train smaller networks which are easier to train.